# Violation of the Mott-Ioffe-Regel Limit: High-temperature Resistivity of Itinerant Magnets $Sr_{n+1}Ru_nO_{3n+1}$ (n=2,3,∞) and $CaRuO_3$

G. Cao, W.H. Song*, Y.P. Sun* and X.N. Lin
Department of Physics and Astronomy, University of Kentucky
Lexington, KY 40506

J. E. Crow
National High Magnetic Field Laboratory, Tallahassee, FL 32306

$Sr_{n+1}Ru_nO_{3n+1}$ represents a class of layered materials whose physical properties are a strong function of the number of Ru-O layers per unit cell, n. This series includes the p-wave superconductor $Sr_2RuO_4$ (n=1), enhanced paramagnetic $Sr_3Ru_2O_7$ (n=2), nearly ferromagnetic $Sr_4Ru_3O_{10}$ (n=3) and itinerant ferromagnetic $SrRuO_3$ (n=∞). In spite of a wide spectrum of physical phenomena, this series of materials along with paramagnetic $CaRuO_3$ shares two major characteristics, namely, robust Fermi liquid behavior at low temperatures and anomalous transport behavior featured by linear temperature dependence of resistivity at high temperature where electron wavepackets are no longer clearly defined. There is no crossover separating such two fundamentally different states. In this paper, we report results of our study that systematically addresses anisotropy and temperature dependence of basal-plane and c-axis resistivity as a function of n for the entire $Sr_{n+1}Ru_nO_{3n+1}$ series and $CaRuO_3$ and for a wide temperature range of 1.7 K<T<900 K. It is found that the anomalous transport behavior correlates with magnetic susceptibility and becomes stronger with decreasing dimensionality. Implications of these results are discussed.

Novel physical phenomena in correlated electron systems present a large array of paradoxes and challenge the very fundamentals of conventional physics such as Fermi liquid theory. A well-known example is the anomalous linear temperature dependence of resistivity, $\rho$, at sufficiently high temperatures in materials such as organic conductors, high temperature superconducting cuprates, and, recently, the p-wave superconductor $Sr_2RuO_4$ [1]. Resistivity saturation at high temperatures, a signature of a breakdown of the Boltzmann theory, is conventionally expected when the mean free path, $l$, of a quasiparticle becomes shorter than the lattice parameter, $a$, or the inverse Fermi wave vector, $k_F^{-1}$, as evidenced in A15-structure superconductors [2, 3]. The requirement for the self-consistent Boltzmann theory of transport leads to the Mott-Ioffe-Regel limit $l \sim a$ or $k_F l \sim 1$ [4,5], or the minimum metallic conductivity $\sigma^{min} \sim 2\pi e^2/3ha$ for a metal with a spherical Fermi surface (similarly, there is the Mooij limit $\rho \sim 100\text{-}150$ $\mu\Omega$ cm [6]), which implies that the concept of a mean free path loses validity and that propagating quasiparticles or electron wavepackets based on Bloch states of definite momentum are no longer clearly defined. In addition, a mechanism based on a two-band model predicts that above a certain temperature the electron-phonon scattering energies may exceed the hybridization interaction, leading to decoupling of the s- and d-states. Conduction then occurs mainly by s electrons with a reduced phonon scattering with increasing temperature, which bring about the decrease in $d\rho/dT$ [7].

The conspicuous lack of both resistivity saturation at high temperatures and a crossover between low and high temperature regions has led to an assertion that the standard theory of Fermi liquids cannot be used to describe these superconducting materials or so-called "bad metals" both below and above the Mott-Ioffe-Regel limit, i.e., the entire temperature range [8]. In spite of the common occurrence of the anomalous behavior, the limited availability of suitable materials has so far allowed only a handful of studies addressing the perplexing coexistence of robust Fermi liquid behavior at low temperatures and the bad metal behavior at sufficiently high temperatures where $l \sim a$ and yet no chemical or structural changes occur. With excellent chemical stability at high temperatures and a wide array of physical properties the Ruddlesden-Popper series of metallic $Sr_{n+1}Ru_nO_{3n+1}$ (n=1,2,3,∞) and $CaRuO_3$ offers a rare opportunity to systematically investigate this issue as a function of correlated electron effects, magnetism and dimensionality. As is known, this series comprises of layers of corner sharing Ru-O octahedra where n is equal to the number of Ru-O layers per formula unit; thus, the progression of n from 1 to ∞ represents an increase in dimensionality. More importantly, the non-Fermi liquid behavior is also observed in these materials such as itinerant ferromagnet $SrRuO_3$ [9-16] that, in contrast to the superconducting materials to which almost all research efforts on this issue have so far been limited, is not in the vicinity of quantum critical point.

$Sr_{n+1}Ru_nO_{3n+1}$, besides the p-wave superconductor $Sr_2RuO_4$ (n=1) [17], includes strongly enhanced paramagnetic $Sr_3Ru_2O_7$ (n=2, $T_M$=18 K) [18-20], nearly ferromagnetic $Sr_4Ru_3O_{10}$ (n=3, Tc=105 K for the c-axis) [21, 22] and itinerant ferromagnetic $SrRuO_3$ (n=∞, $T_C$=165 K) [9-16]. The evolution of (ferro)magnetism in $Sr_{n+1}Ru_nO_{3n+1}$ apparently reflects variations in the Fermi surface topography consistent with the progression of n from 1 to ∞. This is clearly illustrated in Fig.1 where magnetization, M, as a function of temperature for $Sr_{n+1}Ru_nO_{3n+1}$ with n=1,2,3, and ∞ is plotted. That quantum oscillations have recently been observed in both $Sr_4Ru_3O_{10}$ [19] and $SrRuO_3$ [23] confirms that these materials are indeed Fermi liquids at low temperatures. In this paper, we report temperature dependence of basal-plane and c-axis resistivity of $Sr_{n+1}Ru_nO_{3n+1}$ with n=1, 2, 3, ∞ for a wide temperature range of 1.7 K<T<900 K. High temperature resistivity of $Sr_2RuO_4$ (n=1) has been thoroughly studied [1], our data for the n=1 compound only serves as a comparison with that of its sister compounds. All results of our study suggest that the anomalous transport behavior correlates with magnetic susceptibility and becomes stronger with decreasing dimensionality.

The crystals of the entire series of $Sr_{n+1}Ru_nO_{3n+1}$ and $CaRuO_3$ were grown using both floating zone (for n=1, 2) and flux (for n=2, 3, ∞ and $CaRuO_3$) techniques. All crystals studied were characterized by single crystal or powder x-ray diffraction, EDS and TEM, and no impurities and intergrowth were found. The fact that quantum oscillations are observed in $SrRuO_3$ and $Sr_4Ru_3O_{10}$ [22] confirms the high quality of the crystals studied. The Dingle temperature $T_D$ estimated from quantum oscillations, a measure of scattering rate, is in a range of $T_D = 3.0 \pm 0.8$ K [22], comparable to those of good organic metals, whose $T_D$ varies from 0.5 to 3.5 K, indicating the high purity of the single crystals. The high temperature resistivity was measured using a new Displex closed cycle cryostat (Advanced Research Systems DE202) capable for a continuous temperature ramping from 9 K to 900 K. EPOXY E3084, EPOXY T6081 and AREMCO Ceramabond 865 were used as glues for high temperature electrical and thermal contacts, respectively. Low temperature resistivity was done using a function of transport measurements added to a Quantum Design MPMS LX 7T. For each compound, a few crystals were measured. All major features including the magnitude and temperature dependence were reproducible. For compounds with highly anisotropic resistivity, additional efforts were made to assure the rsesistivity that is independent of sample thickness.

A. SrRuO$_3$ and CaRuO$_3$ (n=∞)

All SrRuO$_3$ crystals studied show a sharp Curie temperature at Tc=165 K and a saturation moment of 1.1 μ$_B$/Ru aligned within the basal plane [15]. Shown in Fig. 2 is the resistivity for both the basal plane $\rho_{ab}$ and the c-axis $\rho_c$ as a function of temperature for SrRuO$_3$. The sharp break in slope, corresponding to the Curie temperature, indicates the Fisher-Langer behavior due to short-rang spin fluctuations near Tc, which is recently verified by a critical phenomena study on the system [16]. The resistivity ratio (RRR=$\rho(300K)/\rho(2K)$) ranges from 80-120 and residual resistivity is of 2 μΩ cm for the basal plane for all SrRuO$_3$ crystals studied. It is remarkable that the anisotropy $\rho_c/\rho_{ab}$ is of 2.5 over the entire temperature range measured in spite of a perovskite structure with a mildly orthorhombic distortion (a=5.5636 Å, b=5.5206 Å, c=7.8429 Å). Since resistivity $\rho$ is inversely proportional to $\Omega_p^2\tau$, where $\Omega_p$ is the plasma frequency, $1/\tau$ the scattering rate, this anisotropy implies that $\Omega_p^2\tau$ for the basal plane and the c-axis is significantly different over the broad temperature range. It appears that both $\Omega_p$ and $\tau$ are weak functions of temperature as $\rho_c/\rho_{ab}$ remains essentially a constant throughout the entire temperature range. This feature, which is also seen in Sr$_2$RuO$_4$ below T<20 K [1], is consistent with the prediction of band theory calculations.

SrRuO$_3$ is apparently a Fermi liquid at low temperatures as is evidenced in Fig. 3 where a representative curve of quantum oscillations of magnetization or the de Haas van Alphen effect vs. 1/B is shown for T=30 mK. This effect measures the area of orbits defined by electrons at Fermi surface in magnetic field, B. Although detailed results and analyses will be published elsewhere, it is clear that there are several frequencies involved ranging from 824 T to 1204 T corresponding to different sheets of Fermi surface. In fact, the Fermi liquid behavior is persistent up to T=38 K as $\rho_{ab}$ for the basal plane and $\rho_c$ for the c-axis can be described by $\rho=\rho_o+AT^2$ with $\rho_o$=2.11 $\mu\Omega$ cm and A=9.34x10$^{-3}$ ($\mu\Omega$ cm/K$^2$) for $\rho_{ab}$ (see the inset), for example. Although with slightly different values of A, the quadratic temperature dependence has been previously reported [11, 15, 24]. It appears that the Fermi liquid can only survive at T«0.2$\Theta_D$, where $\Theta_D$ =390 K, a Debye temperature estimated from heat capacity [15]. Noticeably, $\rho \propto T^{3/2}$ for the temperature range of 60<T<130 K. This behavior is reminiscent of that in dilute ferromagnetic alloys where incoherent scattering due to the breakdown of wave vector conservation gives rise to the $T^{3/2}$-dependence for T<Tc [25].

Both $\rho_{ab}$ and $\rho_c$ at T>Tc rise almost linearly with temperature with different slopes and pass through the Mott-Ioffe-Regel limit without a sign of saturation. Fitting the data to $\rho=\rho_o'+\alpha T$ yields $\alpha$ being 0.25 ($\mu\Omega$ cm/K) and 0.65 ($\mu\Omega$ cm/K) for the ab-plane and the c-axis, respectively ($\rho_o'$ is a constant different from $\rho_o$ which depends on elastic scattering). It is noted that $\rho_{ab}$ appears to be quite similar to the resistivity reported in ref. 11. However, no measurable anisotropy was discerned in that work. The lack of the anisotropy was attributed to either a random distribution of orthorhombic domains or twins in the samples [11]. In contrast, $\rho_c$ presented here rises much more rapidly with temperature than $\rho_{ab}$. Using $\rho_{ab}$(300K)=0.15 m$\Omega$ cm and $\rho_c$(300 K)=0.37 m$\Omega$ cm and $\rho^{-1}=(e/12\pi^2 h)$ $(2A_{FS}l)$, where $A_{FS}$ (=1.04 x10$^{17}$ cm$^{-2}$) is the Fermi surface for the paramagnetic state calculated in ref. 11, the upper limit $l$ is estimated to be 10 Å and 2.4 Å for the basal plane and c-axis, comparable to the lattice parameter. While $\rho_{ab}$(300K) may be still in the vicinity of the Mott-Ioffe- Regal limit, $\rho_c$(300K) is clearly beyond the limit, and yet both $\rho_{ab}$ and $\rho_c$ continue to rise linearly with temperature up to 800 K without a visible breakdown of Boltzmann theory.

There have been theoretical studies addressing transport properties of $SrRuO_3$ [12,26, for example]. It is believed that the anomalous behavior may arise from strong p-d electron hybridization that dominates scattering at high temperatures and but becomes less significant at sufficiently low temperatures, thus the recovered Fermi liquid at T<40 K [26]. The anomalous behavior is also attributed to strong electron-paramagnon and even stronger electron-magnon coupling [12].

Indeed, should the electron-magnon coupling play no significant role at T>Tc, the high temperature dependence of resistivity of the paramagnetic metal $CaRuO_3$ [10, 15, 27, 28], which shares a similar electronic structure with $SrRuO_3$ [12], would be comparable. Shown in Fig. 4 is the resistivity of both $SrRuO_3$ and $CaRuO_3$ for the ab-plane (Fig.4a) and the c-axis (Fig.4b). It is apparent that the resistivity for $SrRuO_3$ and $CaRuO_3$ behaves quite differently although $\rho_{ab}$ and $\rho_c$ of $CaRuO_3$ also obey $\rho=\rho_o+AT^2$ at T<40 K with larger $\rho_{oab}$ (=14 $\mu\Omega$ cm) and $A_{ab}$(=45x10$^{-3}$ $\mu\Omega$ cm/K$^2$). (The difference was less significant in our early work where the resistivity was measured only for T<300 K and the ab-plane [15]). For T<Tc, the difference could be largely attributed to the electron-magnon scattering, and reduction of spin scattering which leads to smaller $\rho_{ab}$ and $\rho_c$ below 60 K and 90 K, respectively, in $SrRuO_3$. What is intriguing is that at T>Tc both the magnitude and slope of high temperature resistivity of $CaRuO_3$ is considerably smaller than those of $SrRuO_3$. In addition, the anisotropy $\rho_c/\rho_{ab}$ varies from 1.7 at 5 K to 3.6 at 850 K, also different from that of $SrRuO_3$, which is essentially temperature independent.

Fitting $\rho_{ab}$ of CaRuO$_3$ into $\rho=\rho_o'+\alpha T$ for 290<T<850 K results in $\alpha$ being 0.085 $\mu\Omega$ cm/K, significantly smaller than 0.25 $\mu\Omega$ cm/K for SrRuO$_3$ just mentioned above. If one assumes that the resistivity above Tc is composed of a sum of electron-phonon scattering and electron-paramagnon scattering for both SrRuO$_3$ and CaRuO$_3$, then similar resistivity behavior should be anticipated as a result of the likeness in the electronic structures [12] which are expected to yield similar electron-phonon and electron-paramagnon coupling constants. The distinct behavior shown in Fig. 4a and b implies the existence of additional scattering also responsible for the high temperature resistivity in SrRuO$_3$. It is conceivable that the strong electron-magnon coupling, which may be closely related to the large electron-paramagnon coupling, plays a critical role as well at high temperatures, as predicted in ref. 12. Assuming that $\rho_{CaRuO3}$ for paramagnetic CaRuO$_3$ is a sum of the electron-phonon scattering and electron-paramagnon scattering, then $\Delta\rho$, defined as $\rho_{SrRuO3}-\rho_{CaRuO3}$, should be a fair estimate of the contribution of the electron-magnon scattering to $\rho_{SrRuO3}$ although all corresponding degrees of freedom may be strongly coupled, and it may not be possible to completely separate the electron-magnon scattering from other sources of scattering. $\Delta\rho$, as shown in Fig.4 (dashed line), reveals a few interesting features. First, the magnon related scattering accounts for a major portion of scattering with an even stronger impact on $\rho_{ab}$ than on $\rho_c$. Secondly, $\Delta\rho$ rises with temperature more linearly at T>Tc than both $\rho_{SrRuO3}$ and $\rho_{CaRuO3}$. The slope of the linear T-dependence is 0.19 $\mu\Omega$ cm/K and 0.28 $\mu\Omega$ cm/K for the ab-plane and the c-axis. In addition, as shown in Fig.4c, the magnon related scattering becomes much more anisotropic at T>Tc than at T<Tc. That the paramagnetic CaRuO$_3$, in spite of its more severe orthorhombic distortion, is more conductive than SrRuO$_3$ at high temperatures also suggests significant contributions of magnons and spin fluctuations to the inelastic scattering in the ferromagnetic SrRuO$_3$. This issue is further discussed along with n=1, 2, and 3 compounds below.

B. Sr$_4$Ru$_3$O$_{10}$ (n=3)

With n=3 intermediate between n=2 and ∞, the triple-layered Sr$_4$Ru$_3$O$_{10}$ is characterized by a sharp metamagnetic transition and ferromagnetic behavior occurring within the basal plane and along the c-axis, respectively [22]. Magnetization for the c-axis shows a saturation moment of 1.1 $\mu_B$/Ru and a Curie temperature of Tc=105 K which is then followed by a sharp transition at T$_M$=50 K, [21, 22]. However, M for the basal plane exhibits only a weak cusp at T$_C$, but a pronounced peak at T$_M$, resembling antiferromagnetic-like behavior. Results of our recent study suggest that Sr$_4$Ru$_3$O$_{10}$ is poised between the itinerant metamagnetic state and the itinerant ferromagnetic state that characterize its closest neighbors Sr$_3$Ru$_2$O$_7$ and SrRuO$_3$, respectively [22]

. The anisotropic and complex magnetism is accompanied by critical fluctuations and equally anisotropic and complex transport behavior including large tunneling magnetoresistance and low frequency quantum oscillations [22]. The complex behavior points to an unusual state that shows a delicate balance between fluctuations and order.

Shown in Fig.5 is resistivity of Sr$_4$Ru$_3$O$_{10}$ (n=3) as a function of temperature for the ab-plane and the c-axis for 1.7 K<T<800 K. At low temperatures, Fermi liquid behavior survives as both $\rho_c$ and $\rho_{ab}$ obey $\rho = \rho_o + AT^2$ for the significant regime 1.7-17 K, with $\rho_{oc}$ = 1.30 x 10$^{-3}$ Ω cm, $\rho_{oab}$ = 5.4 x 10$^{-5}$ Ω cm, A$_c$ = 1.04 x 10$^{-5}$ Ω cm/K$^2$ and A$_{ab}$ = 3.4 x 10$^{-7}$ Ω cm/K$^2$ (see the inset). Remarkably, the c-axis values are close to those of Sr$_3$Ru$_2$O$_7$ [18], but the basal plane values are about an order larger; i.e. while triple layer

Sr$_4$Ru$_3$O$_{10}$ has more isotropic resistivity than double layer Sr$_3$Ru$_2$O$_7$ (discussed below), it is not because of better interplanar transport but because of more basal plane scattering. It is also surprising that the anisotropies in A (A$_c$/A$_{ab}$~ 31) and $\rho_0$ ($\rho_{oc}$/$\rho_{oab}$ ~ 24) are similar, since the latter depends only on the band mass and elastic scattering rate, while A depends on the interacting quasiparticle effective mass and inelastic rate. In any case, all these parameters clearly suggest that the Fermi surface remains anisotropic in this triple layer material [22]. Both $\rho_c$ and $\rho_{ab}$ show neither $T^{3/2}$- nor $T^{5/3}$-dependence in the vicinity of the magnetic anomalies expected for a 3D antiferromagnet and ferromagnet, respectively [22]. Beyond the transition regions, $\rho_{ab}$ shows essentially linear temperature dependence for temperature ranges of 18-38 K, 50-100 K and 140-780 K.

The Shubnikov-de Haas effect reveals a very low frequency of $F = 123 \pm 5$ T, which, based on the crystallographic data of Sr$_4$Ru$_3$O$_{10}$ and the Onsager relation $F_0 = A(h/4\pi^2 e)$ ($e$ is the electron charge), corresponds to an area of only 0.9% of the first Brillouin zone [22]. This combined with unusually large $\rho_{ab}$(300 K)=1.2 m$\Omega$ cm implies that $l<a$ and $1/\rho_{ab}<\sigma^{min}$ (~1355 $\Omega^{-1}$ cm$^{-1}$ estimated using the lattice parameters a=5.528 Å, b=5.526 Å, c=28.652 Å [21]), thus precludes metallic Bloch-like behavior at T>300 K. Like resistivity for SrRuO$_3$, $\rho_{ab}$ for 290<T<800 can be described by $\rho=\rho_o'+\alpha T$ with $\alpha$= 1.6 $\mu$W cm/K, an order of magnitude large than that of $\rho_{ab}$ for SrRuO$_3$.

The interplanar resistivity $\rho_c$ shown in Fig.5 also shows various anomalous features fundamentally incompatible with a Fermi liquid description. $\rho_c$ exhibits anomalies corresponding to T$_C$ and T$_M$ whereas $\rho_{ab}$ shows weaker ones at T$_C$ and T$_M$ (not obvious in the figures). The anisotropy $\rho_c/\rho_{ab}$ is unexpectedly large, ranging from nearly 30 at 2 K to 5 at 780 K. The temperature dependent anisotropy suggests temperature dependent $\Omega_p$ and $\tau$, inconsistent with band theory calculations, which predict a temperature independent $\rho_c/\rho_{ab}$ [29]. $\rho_c$ precipitously drops by nearly an order of magnitude from 50 K to 1.7 K. Such a drop in $\rho_c$ is somewhat similar to that of $Sr_2RuO_4$ and $Sr_3Ru_2O_7$ but more pronounced and less anticipated because the triple-layered structure should be more energetically favorable for interplane hopping. This drop is attributed to a drastic reduction of spin-scattering as the system becomes more spin-polarized below $T_M$, which decreases with increasing magnetic field B and eventually vanishes at B=4 T [22]. Remarkably, there is a broad maximum in the vicinity of 300 K signaling a crossover from the metallic state at low temperatures to a nonmetallic state at high temperatures (The brief nonmetallic behavior seen between $T_M$(=50K) and 70 K may be associated with the elongated $RuO_6$ octahedra in the outer layers at low temperatures [21, 22]). This behavior appears to be pertinent to incoherent transport between weakly coupled two-dimensional (2D) non-Fermi liquids, a recent paradigm describing 2D non-Fermi liquids evidenced in a large number of 2D systems such as the cuprates and organic conductors [29]. It is believed that there is confined coherence in metals with sufficiently high anisotropy and strong correlations [29]. The coherent conduction is confined to the basal-plane, and the incoherent interplanar transport occurs, leading to nonmetallic behavior when electrons in the basal-plane become 2D non-Fermi liquids [29]. The c-axis nonmetallic transport at T>300 K presented here is then likely a consequence of incoherent interplanar hopping. The coexistence of incoherent non-metallic transport with coherent metallic transport is a characteristic of 2D non-Fermi liquids, it is striking that it happens in the triplayered $Sr_4Ru_3O_{10}$ in spite of its quasi 3D crystal structure.

### C. $Sr_3Ru_2O_7$ (n=2)

Shown in Fig. 6 is temperature dependence of resistivity for $1.7 \leq T < 850$ K for $Sr_3Ru_2O_7$ (n=2). The bilayered $Sr_3Ru_2O_7$ is an enhanced paramagnet [18] with behavior consistent with proximity to a metamagnetic quantum critical point [19, 20]. Although magnetically isotropic [18], it demonstrates considerably anisotropic transport behavior with $\rho_c/\rho_{ab}$ ranging from 230 at 10 K to 33 at 780 K, a quasi-two dimensional feature consistent with results for $T \leq 300$ K reported earlier [18]. Like $SrRuO_3$ and $Sr_4Ru_3O_{10}$, $Sr_3Ru_2O_7$ also features the $T^2$-dependence of resistivity below 13 K, a characteristic of a Fermi liquid, although the $Sr_3Ru_2O_7$ is the only member in the entire $Sr_{n+1}Ru_nO_{3n+1}$ (n=1,2,3, ∞) series where no quantum oscillations have been discerned or reported yet. While showing some low temperature features common in $Sr_2RuO_4$ and $Sr_4Ru3O_{10}$, this system reveals distinct behavior at high temperatures. At 644 K, $\rho_c$ undergoes an unexpected sharp transition whereas $\rho_{ab}$ does not (see the inset as well). (This feature was reproducible in few other crystals). As seen in the inset, the slope of $\rho_c$ below and above 644 K is clearly different, suggesting a change in scattering rate due to the transition. The transition is more likely to be driven structurally than magnetically, and the origin of it is subject to more investigations. In addition, the temperature dependence of $\rho_{ab}$ at high temperatures (150-780 K) is not quite as linear as that of the other compounds although it becomes almost linear at T>400 K.

Displayed in Fig.7 is a comparison of $\rho_{ab}$ (Fig.7a) and $\rho_c$ (Fig.7b) for $Sr_{n+1}Ru_nO_{3n+1}$ with n=1,2,3, ∞. While the ground state of theses systems is evidently consistent with a Fermi liquid, the high temperature transport clearly violates the Mott-Ioffe-Regel limits, showing not only no saturation in the basal plane rsesitivity but also non-metallic behavior in the c-axis transport for n=1,2, and 3. The sharp contrast between the metallic basal plane and much less or non-metallic interplanar transport at high temperatures underlines the anisotropy which progresses with decreasing the number of $RuO_6$ layers n with n=3 being an exception. It is interesting that $\rho_c$ for all but n=∞ shows a rapid drop below a certain temperature, particularly, for n=1 and 3 (~130 K for n=1, ~50 K for n=3). The origin of it for n=1 and 2 may be associated with suppression of thermal scattering between quasiparticles and phonon as discussed in ref. 1. It may also be an indication of a crossover to a 3D metallic state as temperature decreases. The drop in $\rho_c$ for n=3 is more magnetically driven as it is highly sensitive to magnetic field [22]. Nevertheless, the overall anosotropic behavior for n=1, 2,3 and ∞ appears to be relevant to the concept of the confined coherence in 2D non-Fermi liquids, between which the interplanar hopping is totally incoherent due to interaction effects [29].

Shown in Fig.8a is the slope $\alpha$ of the high temperature $\rho_{ab}$ for n=1, 2, 3 and ∞ as a function of Pauli magnetic susceptibility $\chi_o$ generated by fitting $\rho_{ab}$ for 400<T<800 K into $\rho=\rho_o'+\alpha T$ and magnetic susceptibility $\chi$ for 200<T<400 K into the Cuire-Weiss law respectively. (The Curie-Weiss law is valid as well for materials with local moments as evidenced in refs. 15, 18, 22, for example). While subject to more investigations, the monotonic correlation between the two parameters seems to suggest that $\alpha$ is inversely proportional to $\chi_o$, which is a measure of the density of states near Fermi surface and spin fluctuations. This correlation is rather counter-intuitive and yet interesting. In addition, $\alpha$ increases with decreasing n whereas the coefficient $A_{ab}$ of the $T^2$-term at low temperatures follows somewhat opposite behavior with n=3 being an obvious exception as shown in Fig.8b where $\alpha$ and $A_{ab}$ is plotted as a function of 1/n. The n dependence of $\alpha$ strongly suggests that the scattering mechanism(s) driving the anomalous transport behavior becomes stronger with reducing dimensionality. Relevantly, $\alpha/A_{ab}$ rises almost linearly with $1/n$ (see Fig.8c), suggesting a much stronger dimensionality dependence of $\alpha$ than that of $A_{ab}$, which depends on the quasiparticle effective mass and inelastic scattering rate. That $Sr_4Ru_3O_{10}$ (n=1) frequently stands out as being anomalous is intriguing, consistent with the exotic state reported recently [22].

The layered ruthenates as a new class of materials present a paradox also existing in other highly correlated electron and anisotropic systems such as high Tc cuprates. All these layered ruthenates at low temperatures are without doubt Fermi liquids evidenced by the quadratic temperature dependence of resistivity and, for the n=1, 3 and $\infty$ compounds, also by the observations of quantum oscillations; However, violation of the Mott-Ioffe-Regel limit for the basal plane resistivity and the non-metallic c-axis transport for n=1,2 and 3 clearly reveal behavior inconsistent with any conventional paradigms. Between these two fundamentally different regimes there is no sign of a crossover separating them. This work further highlights invalidity of Fermi liquid theory for describing the anomalous and yet common transport behavior and calls for new experimental and theoretical approaches. The layered ruthenates with a wide spectrum of novel phenomena, high sensitivity of physical properties to dimensionality and undisturbed stoichiometry provide an ideal and rare opportunity for thorough investigations on this issue.

Acknowledgements: This work was supported by NSF grant DMR-0240813.

*Permanent address: Institute of Solid State Physics, Chinese Academy of Sciences, Hefei 230031, Anhui, P.R. China.

Captions:

Fig.1. Magnetization, M, for as a function of temperature for $Sr_{n+1}Ru_nO_{3n+1}$ with n=1,2,3, ∞. Note that the left scale is for n=1 and 2.

Fig.2. Resistivity for both the basal plane $\rho_{ab}$ and the c-axis $\rho_c$ as a function of temperature for $SrRuO_3$ (n=∞) for a temperature range of 1.7<T<800 K.

Fig.3. Amplitude of the de Haas van Alphen effect as a function 1/B at T=30 mK for $SrRuO_3$. Inset: $\rho_{ab}$ as a function of $T^2$ for T≤38 K.

Fig.4. Resistivity of both $SrRuO_3$ and $CaRuO_3$ for the ab-plane (a) and the c-axis (b). Note that the dash lines are $\Delta\rho$ defined as $\rho_{SrRuO3}-\rho_{CaRuO3}$; $\Delta\rho$ for the basal plane and the c-axis as a function of temperature (c).

Fig.5. Resistivity of $Sr_4Ru_3O_{10}$ (n=3) as a function of temperature for the basal plane and the c-axis for 1.7 K<T<800 K. Inset: $\rho_{ab}$ and $\rho_c$ as a function of $T^2$ for T≤17 K.

Fig.6. Temperature dependence of $\rho_{ab}$ and $\rho_c$ for 1.7≤T<850 K for $Sr_3Ru_2O_7$ (n=2). Inset: Enlarged $\rho_c$ near T=644 K.

Fig.7. A comparison of $\rho_{ab}$ (a) and $\rho_c$ (b) for $Sr_{n+1}Ru_nO_{3n+1}$ with n=1,2,3, ∞. Note that $\rho_{ab}$ for n=3 is plotted on the left scale.

Fig.8. The slope α of the high temperature $\rho_{ab}$ for n=1, 2, 3 and ∞ as a function of Pauli magnetic susceptibility $\chi_o$ (a). α and $A_{ab}$ as a function of 1/n (b); $\alpha/A_{ab}$ as a function of 1/n (c). Note that n (=1, 2, 3, and ∞) represents the number of Ru-O layers per formula cell.

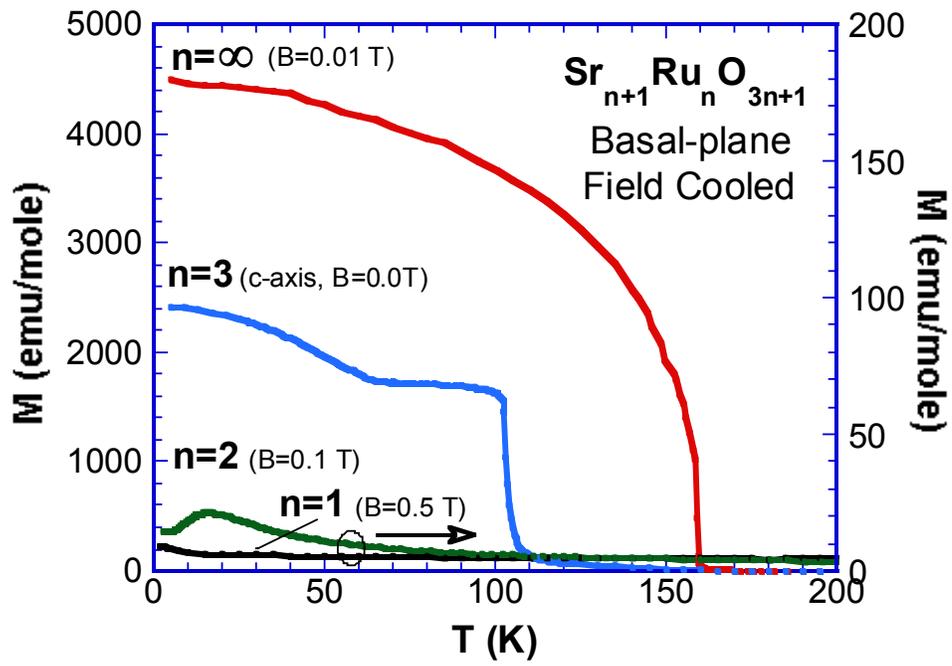

Fig.1, Cao

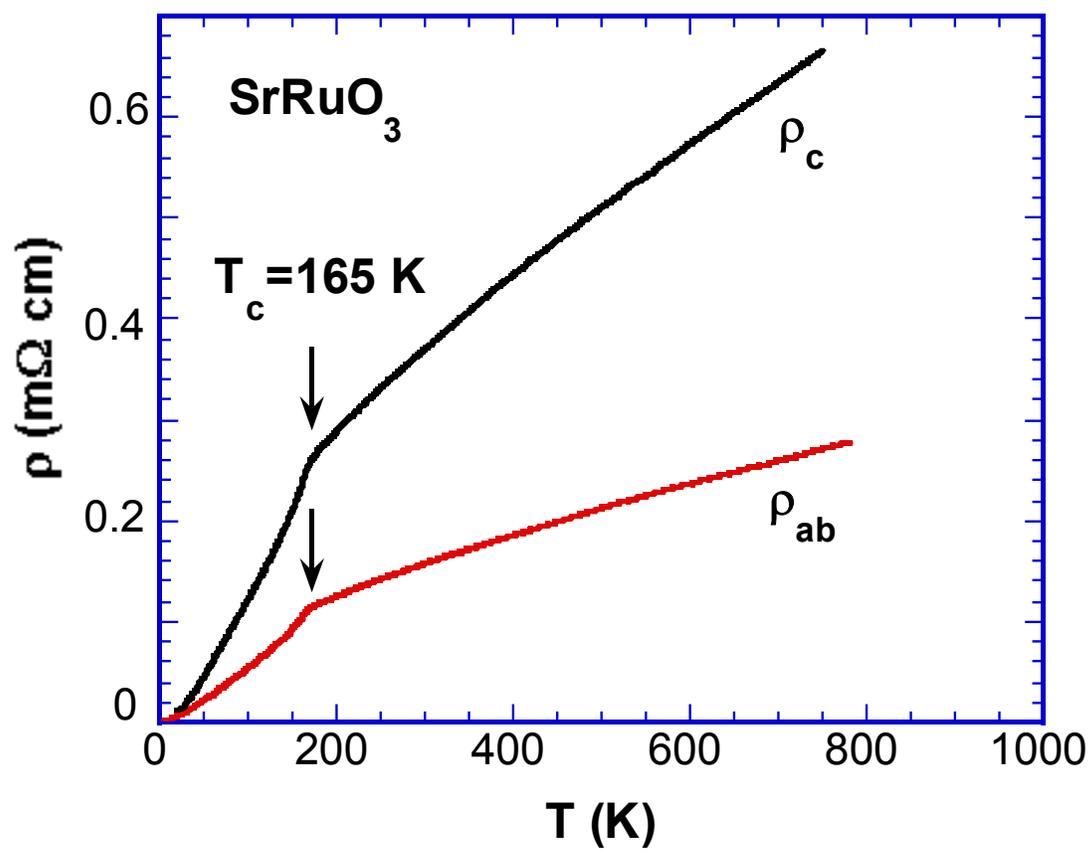

Fig. 2, Cao

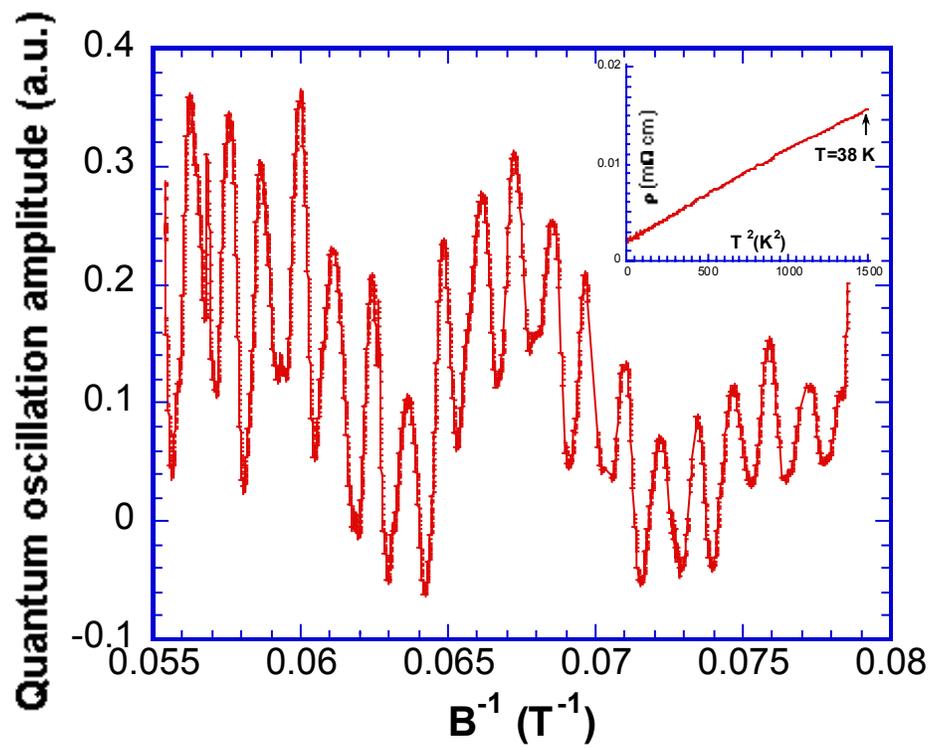

Fig. 3

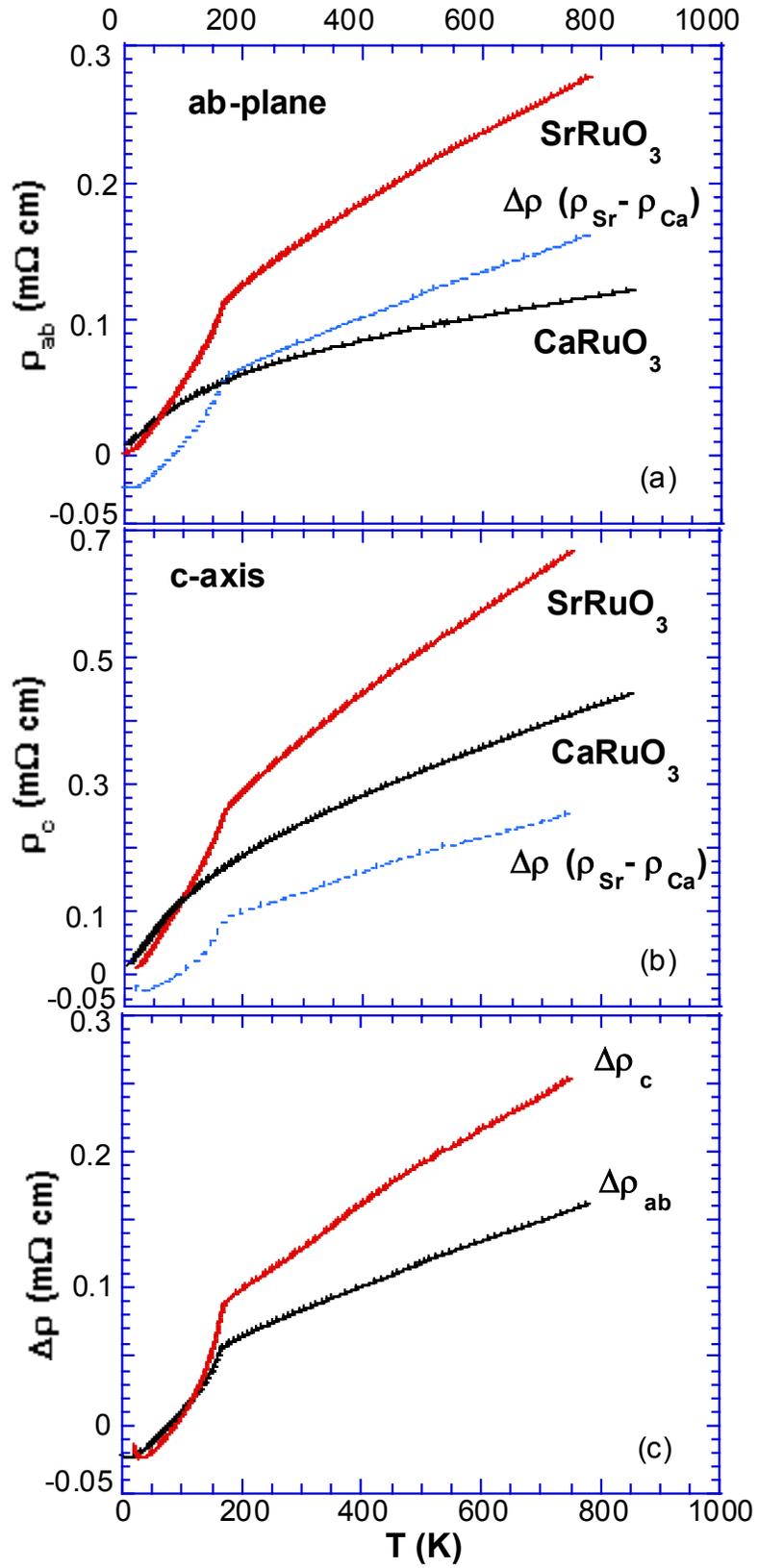

Fig.4

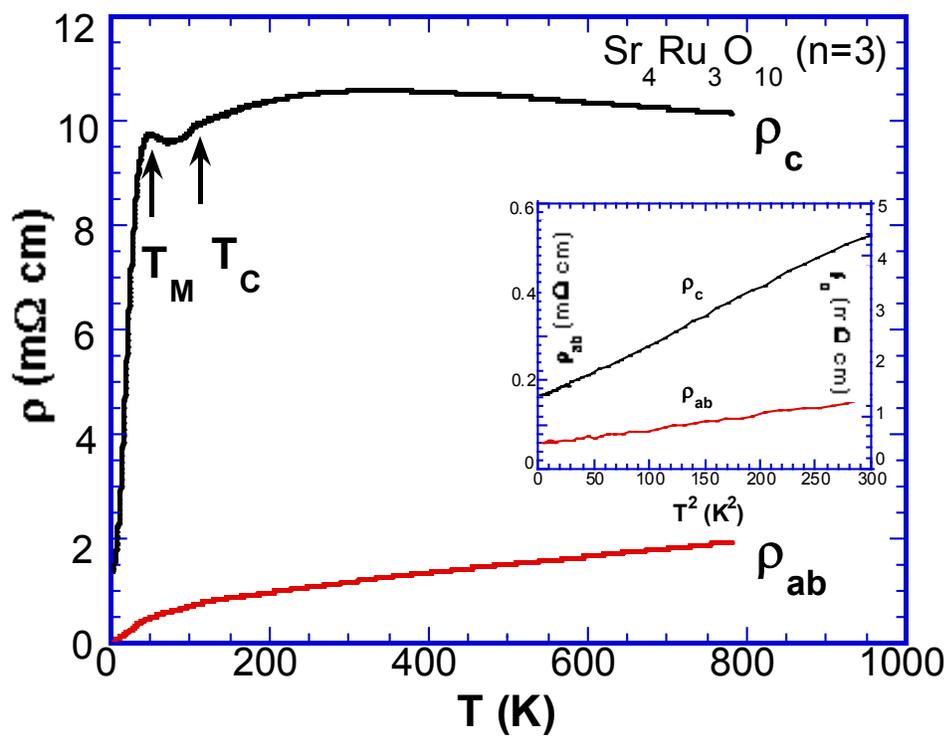

Fig.5

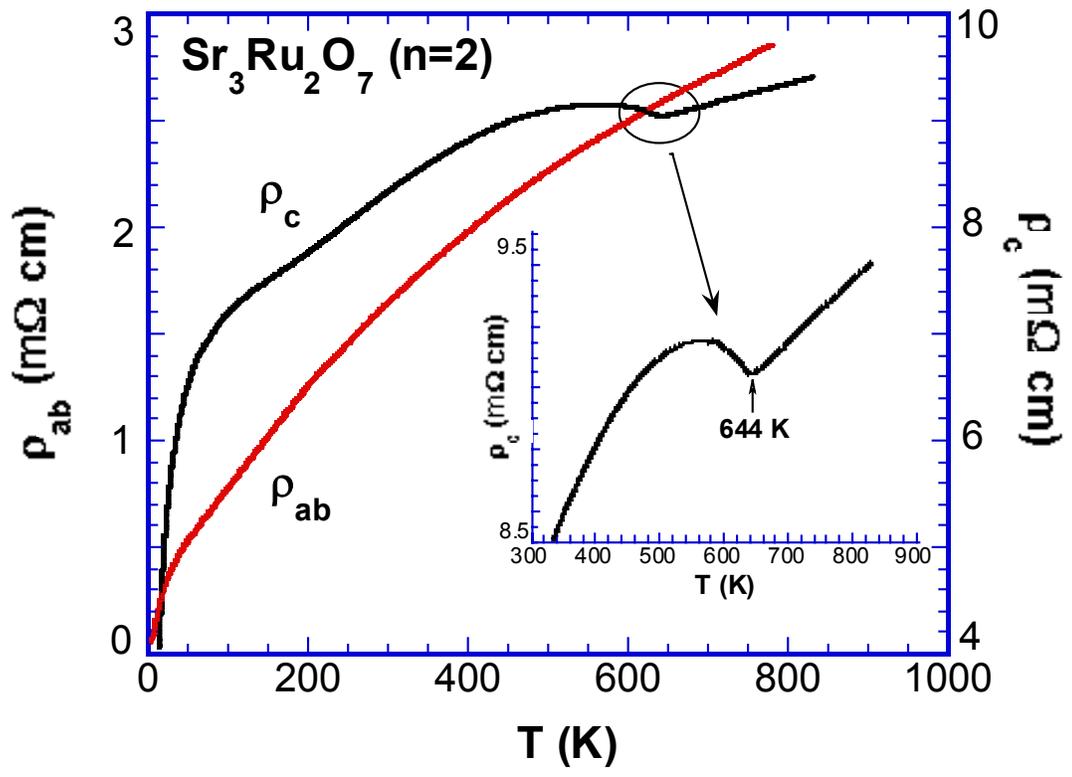

Fig.6

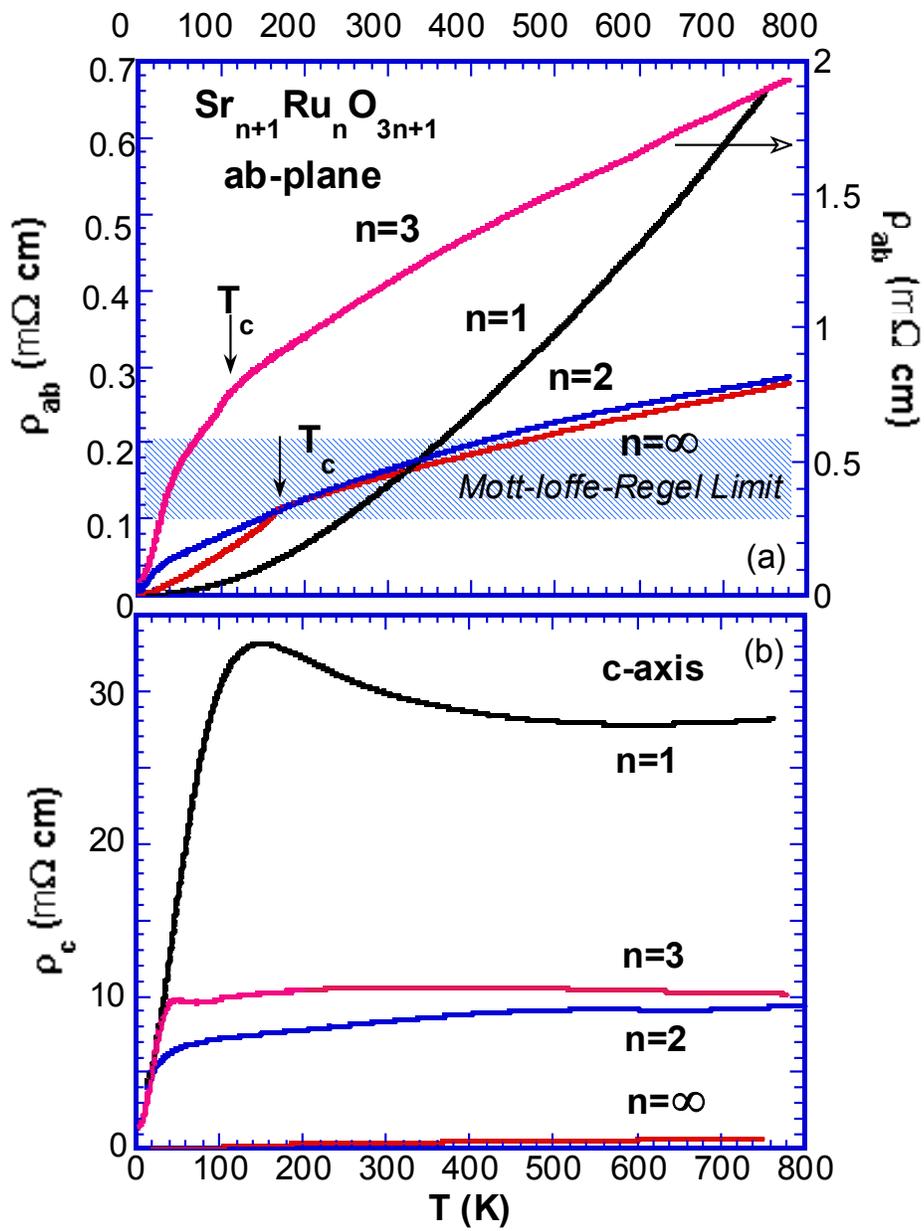

Fig.7

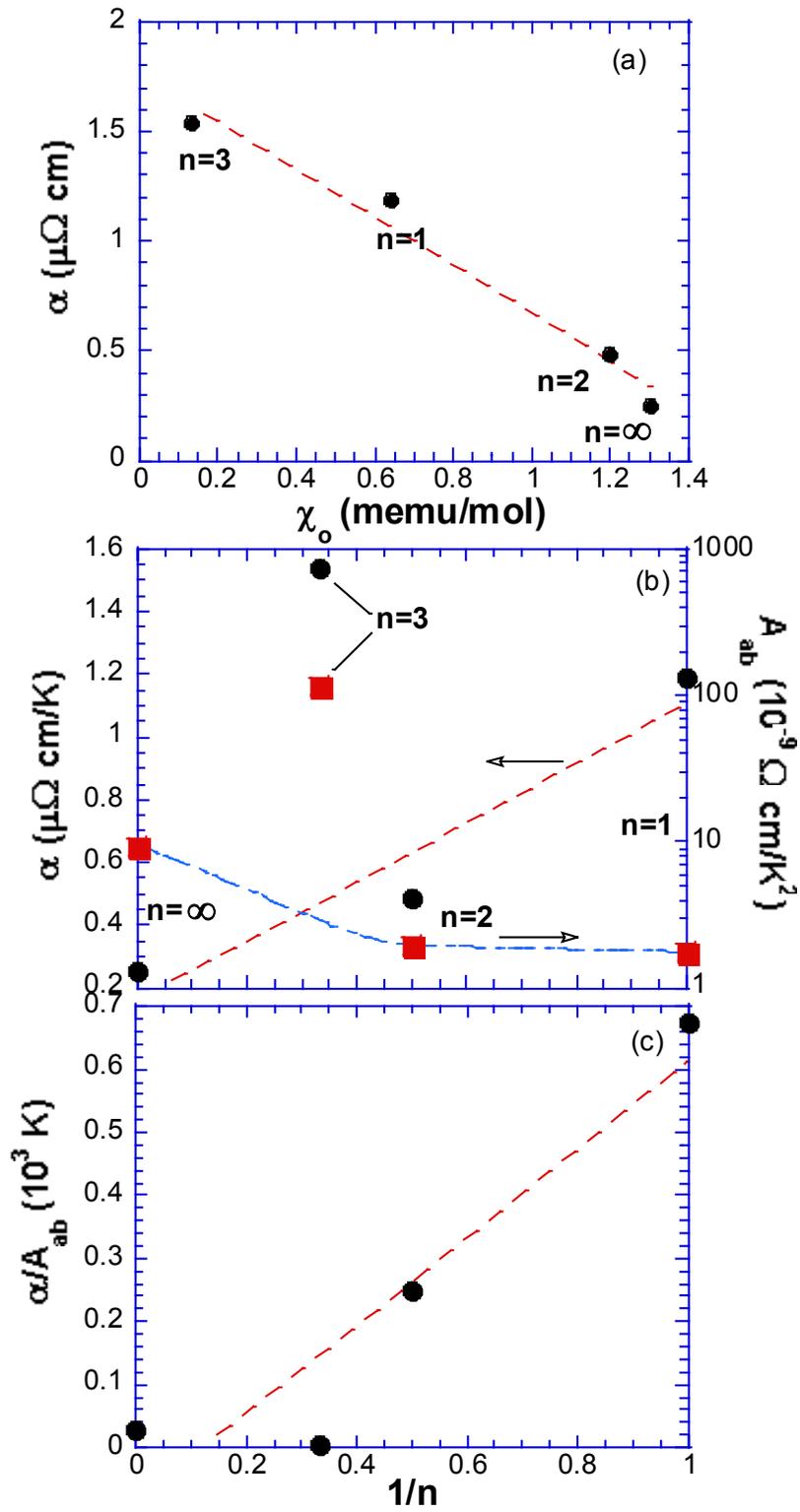

Fig.8